\documentstyle[12pt]{article}

\begin{document}
\begin{titlepage}
\vspace{2.5cm}
\begin{centering}
{\LARGE{\bf Stability analysis of polarized domains}}\\
\bigskip\bigskip
Jos\'e A. Miranda and Michael Widom\\
{\em Department of Physics\\
Carnegie Mellon University\\
Pittsburgh, PA  15213 USA}\\
\vspace{1cm}

\end{centering}
\vspace{1.5cm}

\begin{abstract}

Polarized ferrofluids, lipid monolayers and magnetic bubbles form domains with deformable boundaries. Stability analysis of these domains depends on a family of nontrivial integrals. We present a closed form evaluation of these integrals as a combination of Legendre functions. This result allows exact and explicit formulae for
stability thresholds and growth rates of individual modes. We also evaluate asymptotic behavior in several interesting limits.

\end{abstract}

\begin{center}
\end{center}
\hspace{0.8 cm}
\end{titlepage}
\def\carre{\vbox{\hrule\hbox{\vrule\kern 3pt
\vbox{\kern 3pt\kern 3pt}\kern 3pt\vrule}\hrule}}

\baselineskip = 30pt

Stability analysis predicts the evolution of cylindrical polarized
domains in many systems, including magnetic bubbles~\cite{Thi},
ferrofluid labyrinths~\cite{Lan,Jac} and amphiphilic Langmuir
monolayers~\cite{Gol,McC}. In all these contexts the same family of
nontrivial integrals arises. We evaluate these integrals in closed
form in terms of Legendre functions. Our closed form solution allows
explicit analytic expressions for stability thresholds and individual mode
growth and decay rates. We also present asymptotic behavior of the
integrals in interesting limits such as large aspect ratio, or high
spatial frequency.

Let us briefly describe one particular physical system. Consider an
initially circular (radius $R$) droplet of ferrofluid~\cite{Ros} trapped between two parallel glass plates,
separated by a distance $h$. A convenient measure of the droplet shape is given by the aspect ratio 
\begin{equation}
\label{aspect}
p = {{2R}\over{h}}.
\end{equation}
Apply a uniform magnetic field perpendicular to the plates. Varying the 
magnetic field alters the balance between magnetic interactions and
surface forces. We measure the relative strength of magnetic and
surface forces through the dimensionless magnetic Bond number
\begin{equation}
N_{B}=\frac{2 M^{2} h^{2}}{\gamma},
\end{equation}
where $M$ is the magnetization and $\gamma$ the line tension
(essentially $h$ times the surface tension).  

Jackson, Goldstein and Cebers~\cite{Jac} calculated the growth (or
decay) rate of small time-dependent perturbations $\zeta$ in the
radius $R$ of the initially circular droplet. They model the global fluid flow and energy dissipation using Darcy's law. Let
$\zeta(\phi,t) \sim exp(\sigma_{n}t) cos(n \phi)$, where
$\phi$ represents the polar angle (figure 1). The growth rate $\sigma_{n}$ of
the perturbation by mode $n$ is~\cite{Jac}
\begin{equation}
\label{sigman}
\sigma_{n}=\left (\frac{h \gamma n }{12 \eta R^{3}} \right )  
\left [(1 - n^{2}) + N_{B}{\cal D}_{n}(p) \right ].
\end{equation}
${\cal D}_n(p)$, which depends on mode number $n$ and aspect ratio $p$ but not on the
magnetization, is the complicated integral we will shortly analyze. $\eta$ is the ferrofluid viscosity. An alternate dynamical model~\cite{Lan}, in which energy dissipates locally on the boundary, involves the same integral ${\cal D}_n(p)$.

Now we grapple with the integral. Write 
\begin{equation}
\label{dnp}
{\cal D}_{n}(p)=\left (\frac{1}{x} \right ) \left \{ \left [ I_{n}(0) - I_{1}(0) \right ] 
- \left [ I_{n}(x) - I_{1}(x) \right ] \right \},
\end{equation}
where
\begin{equation}
I_{n}(x)=\int_{0}^{\frac{\pi}{2}}\frac{\sin^{2} n\omega} 
{\sqrt{x + \sin^{2} \omega}} d\omega, \,\,\, x=\frac{1}{p^{2}}.
\end{equation}
This integral is proportional to the magnetic energy when the circular droplet is perturbed by mode $n$. It is derived from a double integral around the perturbed boundary ($\omega=\phi - \phi'$ is the angular separation of points on the boundary). In the late sixties A. A. Thiele~\cite{Thi} derived a number of useful
results on the asymptotic behavior of the integral $I_{n}(x)$ in his study of magnetic bubble stability.
Nevertheless, only a systematic expansion of $I_{n}(x)$ as a function
of the small parameter $x$ has been obtained, an indirect approach
that requires the use of recursion relations and power series
expansions. Even more recent publications [2-5] use a similar approach. It would be useful to find an explicit closed form solution for $I_{n}(x)$.

To evaluate the
integral $I_{n}(x)$ in equation (5), substitute
${\sin^{2} (n \omega)}=\frac{1}{2}[1 - {\cos(2 n \omega)}]$ and change
variables to $\theta=2 \omega$. The utility of the substitution is
evident from inspecting the integral representation for the Legendre
function of the second kind~\cite{Rob}
\begin{equation}
Q_{n - \frac{1}{2}}(z)=\frac{1} {\sqrt{2}} \int_{0}^{\pi}\frac{\cos n\theta} 
{\sqrt{z - \cos \theta}} d\theta, \,\,\,\Re(n)>-\frac{1}{2},\,z>1.
\end{equation}
We find a simple and compact expression for the solution to the integral $I_{n}(x)$,
\begin{equation}
I_{n}(x)=\frac{1} {2} \left [Q_{- \frac{1}{2}}(2x + 1) 
- Q_{n - \frac{1}{2}}(2x + 1) \right ].
\end{equation}

To get closed form expression for ${\cal D}_n(p)$ we need evaluate
the integrals $I_{n}(x)$ at $x=0$. To accomplish this,
we expand $Q_{n - \frac{1}{2}}(2x + 1)$
near $x=0$
\begin{eqnarray}
\label{smallx}
Q_{n - \frac{1}{2}}(2x + 1) & \approx & \left [ -C - \frac{1}{2} 
\log\left (\frac{x}{x + 1} \right ) 
- \psi\left ( n + \frac{1}{2} \right ) \right ] + \nonumber\\
& & \left [ (1 - C) - \frac{1}{2} \log\left (\frac{x}{x + 1} \right ) 
- \psi\left ( n + \frac{1}{2} \right ) \right ] 
\left (\frac{4n^{2} - 1}{4} \right )x \nonumber \\ 
& & + \, {\cal O} (x^{2}),
\end{eqnarray}    
where Euler's psi function $\psi$ is the logarithmic derivative of the Gamma function~\cite{Mag}
and $C$ is Euler's constant. Only the lowest order term contributes to $I_{n}(0)$, but we will need the first order term in our later discussion. Thus
\begin{equation}
I_{n}(0)=\frac{1}{2} \left [\psi(n + \frac{1}{2}) - \psi(\frac{1}{2}) \right ],
\end{equation}
and finally we rewrite equation~(\ref{dnp})
\begin{equation}
\label{finaldnp}
{\cal D}_n(p)=
{\frac{p^2}{2} \left \{ \left [\psi(n + \frac{1}{2}) 
- \psi(- \frac{1}{2}) \right ] + \left [ Q_{n - \frac{1}{2}}(\frac{p^2 + 2}{p^2}) 
- Q_{\frac{1}{2}}(\frac{p^2 + 2}{p^2}) \right ] \right \} }.
\end{equation}
This is our central result, yielding exact explicit dependence of mode growth rates on Bond number $N_{B}$, mode number $n$ and aspect ratio $p$.

Figure 2 illustrates the mode growth rate as a function of magnetic Bond number $N_{B}$ for several mode numbers $n$. We plot the dimensionless growth rate
\begin{equation}
\bar{\sigma}_{n}=\left ( \frac{12 \eta R^{3}}{h \gamma} \right ) \sigma_{n},
\end{equation}
with $\sigma_{n}$ defined as in equation (3). The aspect ratio is held fixed at $p=20$. Note that mode $n=0$ is meaningless because of the constraint of constant volume, while mode $n=1$ is just a spatial translation, so we consider only $n \geq 2$. Mode $n$ goes unstable when its growth rate $\sigma_n$ vanishes,
at critical Bond number
\begin{equation}
\label{nbstar}
N_B^{\star}={{(n^2-1)}\over{{\cal D}_n(p)}}.
\end{equation}
The first mode to go unstable as Bond number increases is $n=2$. However,
if the magnetic Bond number is suddenly increased to a value beyond the $n=2$ threshold, the fastest
growing mode $n^{\star}$ may have a value greater than 2.

In figure 3 we plot the behavior of $N_B^{\star}$ in terms of the ratio $n/p$. This ratio is proportional to the spatial wave vector $k$ of the perturbation. In particular, the wave vector $k=n/R$ and we find 
\begin{equation}
\frac{n}{p}=\frac{kh}{2}. 
\end{equation}
It appears in figure 3 that all curves $(2 \leq n \leq 15)$ converge to a limiting function as the value of $n$ increases. For large $n$ and $p$, $N_B^{\star}$ becomes a function only of the ratio $n/p$. Below we analyze in more detail the asymptotic behavior for large $n$ and $p$.

In many practical situations, the aspect ratio $p$ or the mode number
$n$ assume large values [2-5], so we calculate asymptotic expansions
taking advantage of the closed form expression for ${\cal D}_n(p)$ derived above. First
consider large $n$ and $p$, for fixed ratio $n/p$. Use the asymptotic
limit~\cite{Mag}
\begin{equation}
\lim_{n \rightarrow \infty} Q_{n - \frac{1}{2}}(\cosh \left( \frac{q}{n} \right))=K_{0}(q), 
\end{equation}
where $K_{0}(q)$ is the modified Bessel function of order zero, and set 
\begin{equation}
q=n \cosh^{-1} \left( 1 + \frac{2}{p^{2}} \right ) \approx \frac{2n}{p} + {\cal O} \left( \frac{n}{p^{3}} \right ).  
\end{equation}
Note that
\begin{equation}
\lim_{n \rightarrow \infty} \psi(n + \frac{1}{2}) \approx \log(n),
\end{equation}
to get
\begin{equation}
\label{largen}
{\cal D}_n(p) \approx {\frac{p^2}{2} \left \{ \log \left( \frac{n}{p} \right ) + K_{0} 
\left(\frac{2n}{p} \right) + C \right \} }.
\end{equation}
This result is quite general, being valid for large values of $p$ and $n$, provided $n/p$ is kept finite. A related expression may be found in reference~\cite{Lan}.

From equation (12) for the critical Bond number $N_B^{\star}$ and asymptotic expansion~(\ref{largen}) for ${\cal D}_n(p)$ we find,
\begin{equation}
N_B^{\star} \approx \frac{(\frac{n}{p})^2}{ \frac{1}{2} \left \{ \log \left( \frac{n}{p} \right ) + K_{0} 
\left(\frac{2n}{p} \right) + C \right \} }.
\end{equation}
As expected (see figure 3) $N_B^{\star}$ is a function only of the ratio $n/p$. The Bessel function $K_{0}$ simplifies for small or large arguments.
In particular, the limit of small $n/p$ yields
\begin{equation}
\label{largenp}
N_B^{\star} \approx \frac{1}{ \frac{1}{2} \left \{ 1 - C - \log \left( \frac{n}{p} \right ) \right \} },
\end{equation}
while in the opposite limit of large $n/p$,
\begin{equation}
N_B^{\star} \approx \frac{(\frac{n}{p})^2}{ \frac{1}{2} \left \{ \log \left( \frac{n}{p} \right ) + C  \right \} }.
\end{equation}

Next consider the situation in which $p$ is large, but no assumption
is made regarding the value of $n$. This limit is relevant, for example, to the critical Bond number of mode $n=2$, in the limit of large aspect ratio. Typical values of $p$ range from the optimal value for magnetic bubbles~\cite{Thi2} $p=2$ through a couple of orders of magnitude for ferrofluid thin films~\cite{Bou} and reach as high as $p=10^{4}$ for lipid monolayers~\cite{McC}. The large $p$ limit can be
obtained from equation~(\ref{smallx}), where the Legendre function of
the second kind has been expanded to first order in terms of $x=1/p^{2}$,
resulting in
\begin{equation}
\label{largep}
{\cal D}_n(p) \approx \frac{1}{2} (n^{2} - 1) \left \{ \log(4p) - 1
- \frac{1}{4} \left ( \frac{4n^{2} - 1}{n^{2} - 1} \right )
\left [ \psi(n + \frac{1}{2}) - \psi(-\frac{1}{2}) \right ] \right \},
\end{equation}
valid for arbitrary $n$. The difference of $\psi$ functions in equation (21) 
replaces the infinite series equation (2.12) in reference [4].
Taking the large $n$ limit of~(\ref{largep}) agrees with the small $n/p$ limit of equation~(\ref{largen}).

We conclude by discussing the flat edge limit. Tsebers and Maiorov~\cite{Tse} studied the threshold for instability when a thin layer of a magnetic liquid is placed  in a homogeneous magnetic field perpendicular to the layer. The interface is flat with a small perturbation of wave vector $k$. To get this flat edge limit from our cylindrical geometry, we take radius $R \rightarrow \infty$ at fixed h. Fixed wave vector $k$ implies $n$ and $p$ $\rightarrow \infty$ with fixed $n/p$ as in our asymptotic expansion equations (17) and (18). Therefore, replacing $n/p$ with the wave vector $k$ as in equation (13) we obtain
\begin{equation}
N_B^{\star} \approx \frac{{(kh)}^{2}}{ 2 \left \{ \log \left( \frac{kh}{2} \right ) + K_{0}(kh) + C \right \}}.
\end{equation}
This result agrees with the prediction of Tsebers and Maiorov~\cite{Tse} in zero gravity.

Still considering the flat edge limit cited above, we examine the  expression for the mode growth rate,
\begin{equation}
\sigma(k) \approx \left (\frac{\gamma k }{12 \eta h} \right )  
\left \{ 2 N_{B} \left [ \log \left( \frac{kh}{2} \right ) + K_{0}(kh) + C \right ] - (k h)^2 \right \}.
\end{equation}
Taking the limit in which the wavelengths are large compared to the slab thickness ($kh \ll 1$), we expand the Bessel function $K_{0}$ in equation (23) to the second order in $kh$. In this limit, the critical wave number $k_{c}$ (defined by setting $\sigma(k)=0$) is
\begin{equation}
k_{c} \approx \frac{2}{h} e^{1 - C} e^{- \frac{2}{N_{B}}},
\end{equation}
and the fastest growing mode $k^{\star}$ (defined by setting $d\sigma(k)/dk=0$) is
\begin{equation}
k^{\star} \approx \frac{2}{h} e^{\frac{2}{3} - C} e^{- \frac{2}{N_{B}}}.
\end{equation}
Comparing equations (24) and (25) we see that the ratio of spatial frequencies between the critical and the fastest growing mode is
\begin{equation}
\frac{k_{c}}{k^{\star}}=e^{\frac{1}{3}}.
\end{equation}
This result agrees with McConnell~\cite{McC2}, who used a simple hydrodynamic model to study dynamical stability analysis of a straight edge on a large lipid monolayer domain.

\pagebreak

\noindent
{\bf Acknowledgments}\\
\noindent
We would like to thank A. A. Thiele for useful conversations.
J.A.M. (CNPq reference number 200204/93-9) would like to thank CNPq
(Brazilian Research Council) for financial support. This work was
supported in part by the National Science Foundation grant
DMR-9221596.

\pagebreak

\noindent
{\Large {\bf Figure Captions}}
\vskip 0.5 in
\noindent
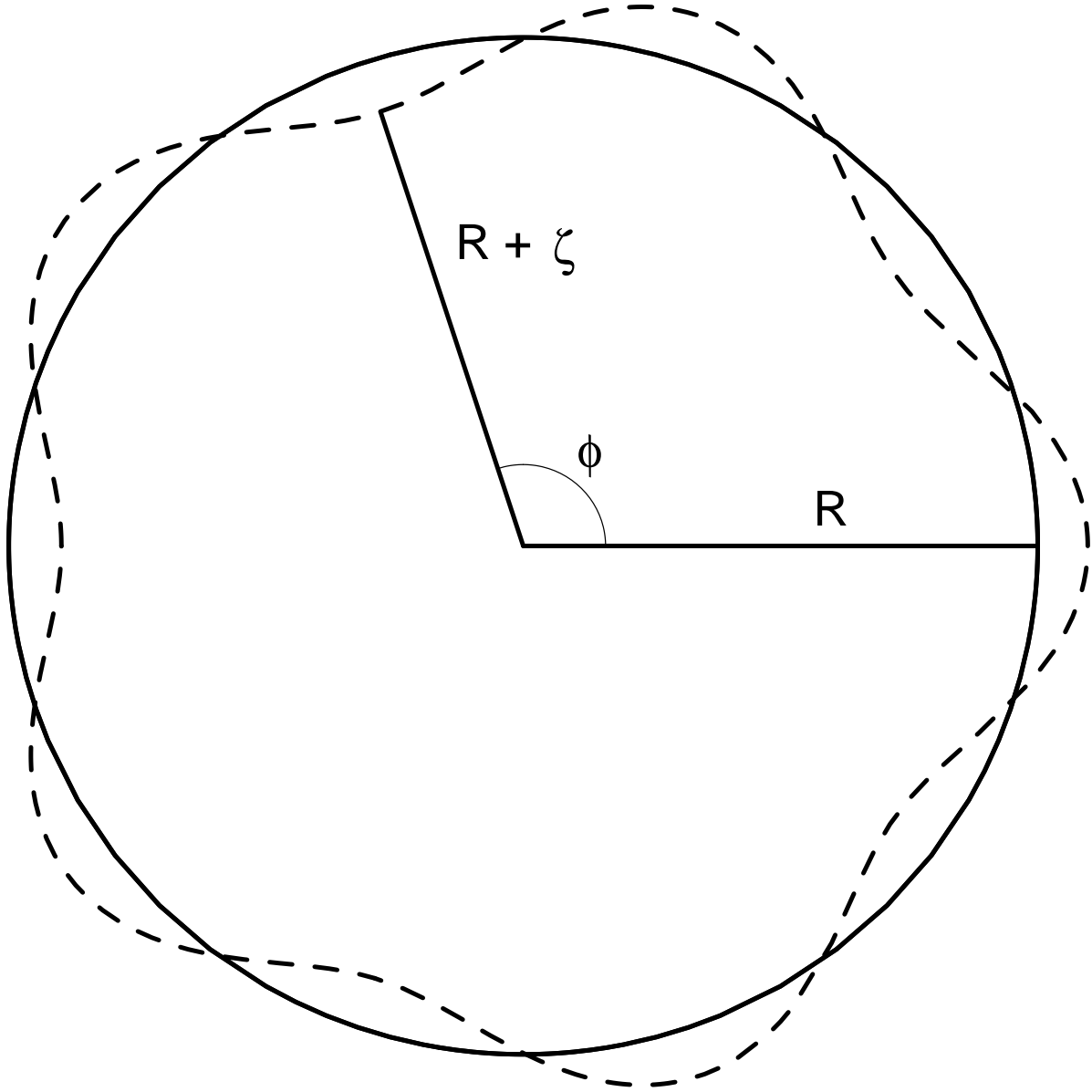
{\bf Figure 1:} Schematic representation of an initially circular ferrofluid droplet of radius $R$ (solid curve) and its perturbation by $\zeta$ for the $n=5$ mode (dashed curve). The angular location of points on the boundary is given by $\phi$. The droplet volume is kept constant, since the fluid is incompressible.
\vskip 0.25 in
\noindent
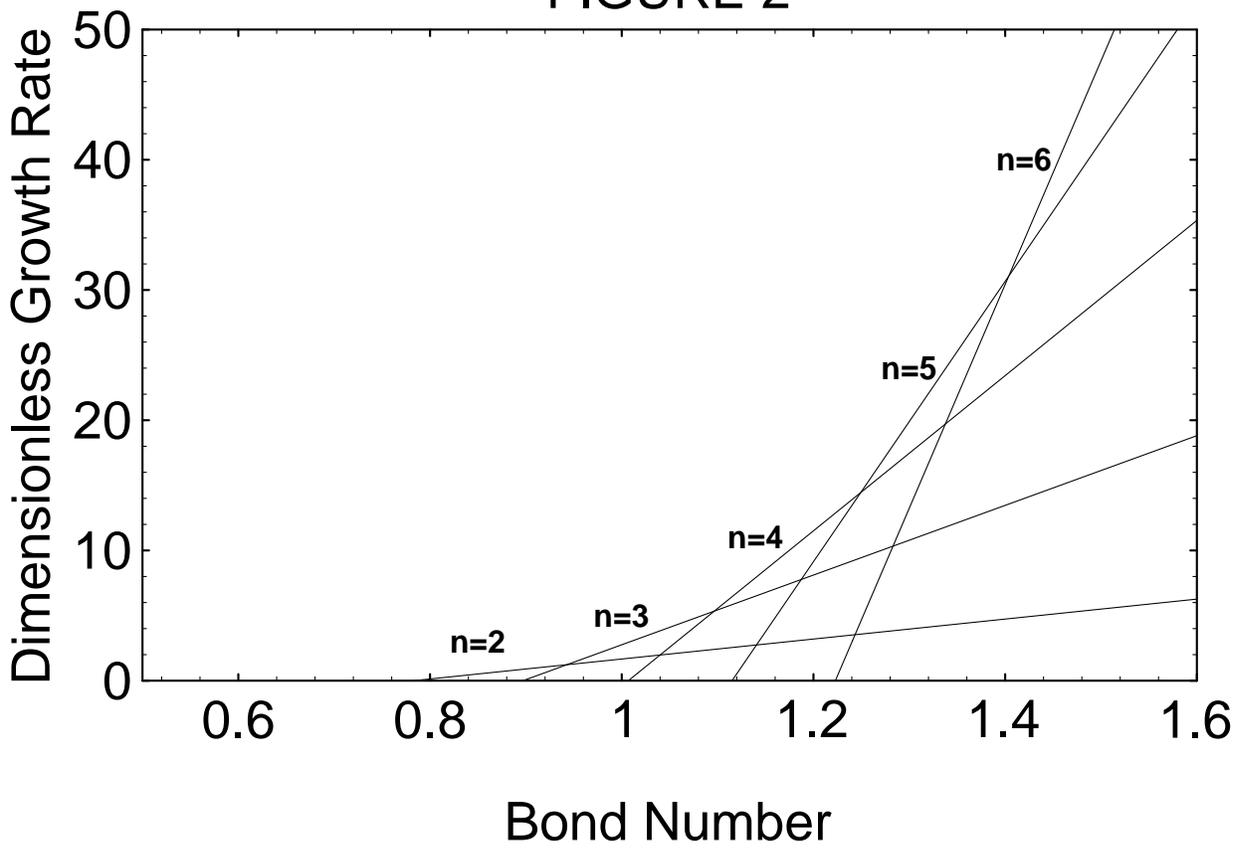
{\bf Figure 2:} Dimensionless growth rate $\bar{\sigma}_{n}$ (see equation (11) in the text), for mode numbers $2 \leq n \leq 6$, as a function of the magnetic Bond number $N_{B}$. The aspect ratio $p=20$. Notice that for each $n$ there exists a critical Bond number (when $\bar{\sigma}_{n}=0$) and for each value of $N_{B}$ there is a fastest growing mode $n^{\star}$.
\vskip 0.25 in
\noindent
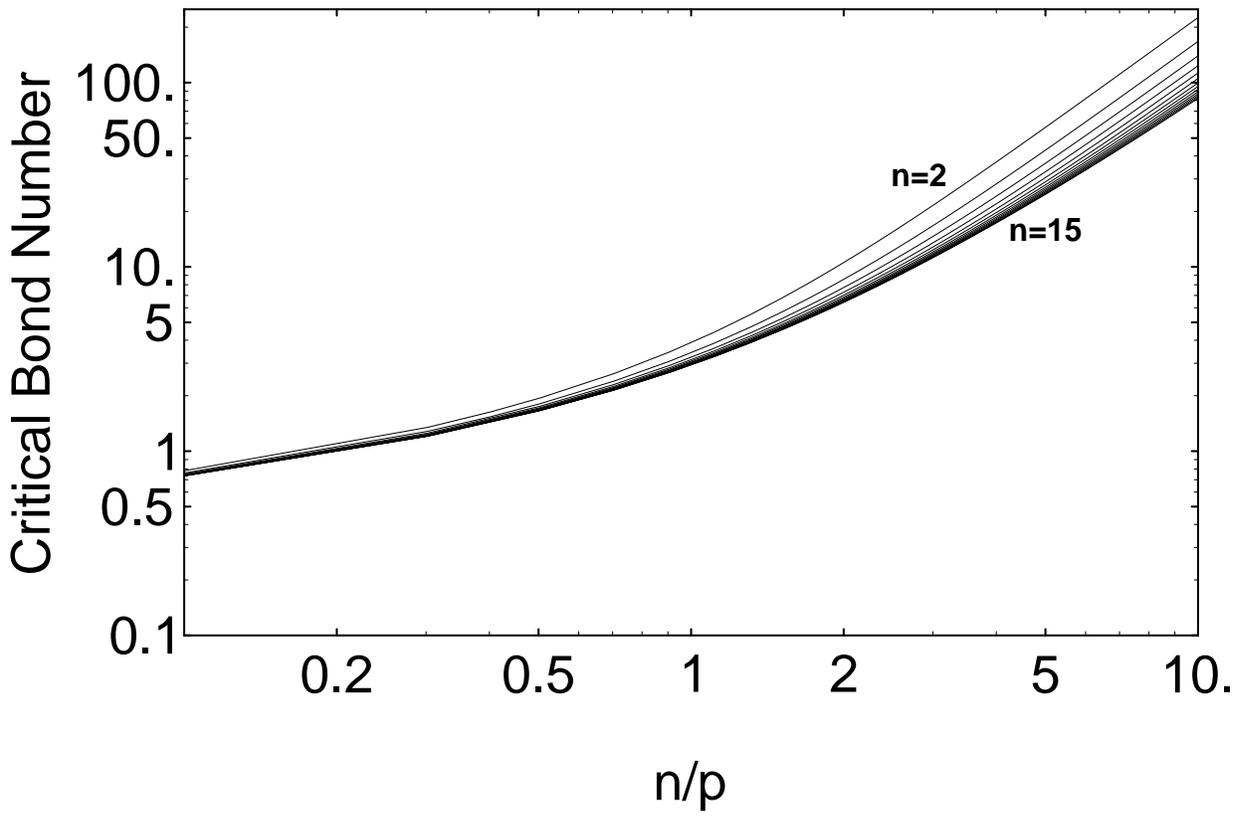
{\bf Figure 3:} Critical Bond number $N_B^{\star}$ for $2 \leq n \leq 15$, as a function of the ratio $n/p$. All curves are obtained exactly, combining the closed form expression in equation (10) with equation (12).

\end{document}